\documentclass[aps,prl,twocolumn,superscriptaddress,showpacs]{revtex4}
\usepackage{graphicx} 
\usepackage{dcolumn}
\usepackage{bm}
\usepackage{setspace}
\usepackage{epsfig}
\usepackage{epstopdf}
\def\met{$p_{T}^{\nu}$}

\def\gammaw{$\Gamma_W$}
\def\mw{$M_W$}
\def\Gf{$G_F$}

\def\ipar{$d_0$}
\def\mt{$M_T$}

\def\ptw{$p_T^W$}
\def\ptm{$p_T^\mu$}

\def\ptlvec{$\vec{p}_{T}^{\,\,\ell}$}
\def\ptnuvec{$\vec{p}_{T}^{\,\,\nu}$}
\begin{document}
\title{\boldmath A Direct Measurement of the $W$ Boson Width in $p\bar{p}$ Collisions at $\sqrt{s}=1.96$ TeV}
\affiliation{Institute of Physics, Academia Sinica, Taipei, Taiwan 11529, Republic of China} 
\affiliation{Argonne National Laboratory, Argonne, Illinois 60439} 
\affiliation{Institut de Fisica d'Altes Energies, Universitat Autonoma de Barcelona, E-08193, Bellaterra (Barcelona), Spain} 
\affiliation{Baylor University, Waco, Texas  76798} 
\affiliation{Istituto Nazionale di Fisica Nucleare, University of Bologna, I-40127 Bologna, Italy} 
\affiliation{Brandeis University, Waltham, Massachusetts 02254} 
\affiliation{University of California, Davis, Davis, California  95616} 
\affiliation{University of California, Los Angeles, Los Angeles, California  90024} 
\affiliation{University of California, San Diego, La Jolla, California  92093} 
\affiliation{University of California, Santa Barbara, Santa Barbara, California 93106} 
\affiliation{Instituto de Fisica de Cantabria, CSIC-University of Cantabria, 39005 Santander, Spain} 
\affiliation{Carnegie Mellon University, Pittsburgh, PA  15213} 
\affiliation{Enrico Fermi Institute, University of Chicago, Chicago, Illinois 60637} 
\affiliation{Comenius University, 842 48 Bratislava, Slovakia; Institute of Experimental Physics, 040 01 Kosice, Slovakia} 
\affiliation{Joint Institute for Nuclear Research, RU-141980 Dubna, Russia} 
\affiliation{Duke University, Durham, North Carolina  27708} 
\affiliation{Fermi National Accelerator Laboratory, Batavia, Illinois 60510} 
\affiliation{University of Florida, Gainesville, Florida  32611} 
\affiliation{Laboratori Nazionali di Frascati, Istituto Nazionale di Fisica Nucleare, I-00044 Frascati, Italy} 
\affiliation{University of Geneva, CH-1211 Geneva 4, Switzerland} 
\affiliation{Glasgow University, Glasgow G12 8QQ, United Kingdom} 
\affiliation{Harvard University, Cambridge, Massachusetts 02138} 
\affiliation{Division of High Energy Physics, Department of Physics, University of Helsinki and Helsinki Institute of Physics, FIN-00014, Helsinki, Finland} 
\affiliation{University of Illinois, Urbana, Illinois 61801} 
\affiliation{The Johns Hopkins University, Baltimore, Maryland 21218} 
\affiliation{Institut f\"{u}r Experimentelle Kernphysik, Universit\"{a}t Karlsruhe, 76128 Karlsruhe, Germany} 
\affiliation{Center for High Energy Physics: Kyungpook National University, Daegu 702-701, Korea; Seoul National University, Seoul 151-742, Korea; Sungkyunkwan University, Suwon 440-746, Korea; Korea Institute of Science and Technology Information, Daejeon, 305-806, Korea; Chonnam National University, Gwangju, 500-757, Korea} 
\affiliation{Ernest Orlando Lawrence Berkeley National Laboratory, Berkeley, California 94720} 
\affiliation{University of Liverpool, Liverpool L69 7ZE, United Kingdom} 
\affiliation{University College London, London WC1E 6BT, United Kingdom} 
\affiliation{Centro de Investigaciones Energeticas Medioambientales y Tecnologicas, E-28040 Madrid, Spain} 
\affiliation{Massachusetts Institute of Technology, Cambridge, Massachusetts  02139} 
\affiliation{Institute of Particle Physics: McGill University, Montr\'{e}al, Canada H3A~2T8; and University of Toronto, Toronto, Canada M5S~1A7} 
\affiliation{University of Michigan, Ann Arbor, Michigan 48109} 
\affiliation{Michigan State University, East Lansing, Michigan  48824} 
\affiliation{University of New Mexico, Albuquerque, New Mexico 87131} 
\affiliation{Northwestern University, Evanston, Illinois  60208} 
\affiliation{The Ohio State University, Columbus, Ohio  43210} 
\affiliation{Okayama University, Okayama 700-8530, Japan} 
\affiliation{Osaka City University, Osaka 588, Japan} 
\affiliation{University of Oxford, Oxford OX1 3RH, United Kingdom} 
\affiliation{University of Padova, Istituto Nazionale di Fisica Nucleare, Sezione di Padova-Trento, I-35131 Padova, Italy} 
\affiliation{LPNHE, Universite Pierre et Marie Curie/IN2P3-CNRS, UMR7585, Paris, F-75252 France} 
\affiliation{University of Pennsylvania, Philadelphia, Pennsylvania 19104} 
\affiliation{Istituto Nazionale di Fisica Nucleare Pisa, Universities of Pisa, Siena and Scuola Normale Superiore, I-56127 Pisa, Italy} 
\affiliation{University of Pittsburgh, Pittsburgh, Pennsylvania 15260} 
\affiliation{Purdue University, West Lafayette, Indiana 47907} 
\affiliation{University of Rochester, Rochester, New York 14627} 
\affiliation{The Rockefeller University, New York, New York 10021} 
\affiliation{Istituto Nazionale di Fisica Nucleare, Sezione di Roma 1, University of Rome ``La Sapienza," I-00185 Roma, Italy} 
\affiliation{Rutgers University, Piscataway, New Jersey 08855} 
\affiliation{Texas A\&M University, College Station, Texas 77843} 
\affiliation{Istituto Nazionale di Fisica Nucleare, University of Trieste/\ Udine, Italy} 
\affiliation{University of Tsukuba, Tsukuba, Ibaraki 305, Japan} 
\affiliation{Tufts University, Medford, Massachusetts 02155} 
\affiliation{Waseda University, Tokyo 169, Japan} 
\affiliation{Wayne State University, Detroit, Michigan  48201} 
\affiliation{University of Wisconsin, Madison, Wisconsin 53706} 
\affiliation{Yale University, New Haven, Connecticut 06520} 
\author{T.~Aaltonen}
\affiliation{Division of High Energy Physics, Department of Physics, University of Helsinki and Helsinki Institute of Physics, FIN-00014, Helsinki, Finland}
\author{J.~Adelman}
\affiliation{Enrico Fermi Institute, University of Chicago, Chicago, Illinois 60637}
\author{T.~Akimoto}
\affiliation{University of Tsukuba, Tsukuba, Ibaraki 305, Japan}
\author{M.G.~Albrow}
\affiliation{Fermi National Accelerator Laboratory, Batavia, Illinois 60510}
\author{B.~\'{A}lvarez~Gonz\'{a}lez}
\affiliation{Instituto de Fisica de Cantabria, CSIC-University of Cantabria, 39005 Santander, Spain}
\author{S.~Amerio}
\affiliation{University of Padova, Istituto Nazionale di Fisica Nucleare, Sezione di Padova-Trento, I-35131 Padova, Italy}
\author{D.~Amidei}
\affiliation{University of Michigan, Ann Arbor, Michigan 48109}
\author{A.~Anastassov}
\affiliation{Rutgers University, Piscataway, New Jersey 08855}
\author{A.~Annovi}
\affiliation{Laboratori Nazionali di Frascati, Istituto Nazionale di Fisica Nucleare, I-00044 Frascati, Italy}
\author{J.~Antos}
\affiliation{Comenius University, 842 48 Bratislava, Slovakia; Institute of Experimental Physics, 040 01 Kosice, Slovakia}
% ...
\collaboration{CDF Collaboration\footnote{With visitors from $^a$University of Athens, 15784 Athens, Greece, 
$^b$Chinese Academy of Sciences, Beijing 100864, China, 
$^c$University of Bristol, Bristol BS8 1TL, UK, 
$^d$University Libre de Bruxelles, B-1050 Brussels, Belgium, 
$^e$University of California Irvine, Irvine, CA  92697, 
$^f$University of California Santa Cruz, Santa Cruz, CA  95064, 
$^g$Cornell University, Ithaca, NY  14853, 
$^h$University of Cyprus, Nicosia CY-1678, Cyprus, 
$^i$University College Dublin, Dublin 4, Ireland, 
$^j$University of Edinburgh, Edinburgh EH9 3JZ, UK, 
$^k$University of Heidelberg, D-69120 Heidelberg, Germany, 
$^l$Universidad Iberoamericana, Mexico D.F., Mexico, 
$^m$University of Manchester, Manchester M13 9PL, UK, 
$^n$Nagasaki Institute of Applied Science, Nagasaki, Japan, 
$^o$University de Oviedo, E-33007 Oviedo, Spain, 
$^p$Queen Mary, University of London, London E1 4NS, UK, 
$^q$Texas Tech University, Lubbock, TX  79409, 
$^r$IFIC(CSIC-Universitat de Valencia), 46071 Valencia, Spain, 
}}
\noaffiliation

\date{\today}
\begin{abstract}
A direct measurement of the total decay width of the $W$ boson
{\gammaw} is presented using  350~pb$^{-1}$ of data from $p\overline{p}$ collisions at
$\sqrt{s} = 1.96$~TeV collected with the CDF II detector at the
Fermilab Tevatron. The width is determined by normalizing predicted
signal and background distributions to 230185 $W$ candidates decaying to
$e\nu$ and $\mu\nu$ in the
transverse-mass region $50 < M_T < 90$~GeV and then fitting the predicted 
shape to 6055 events in the high-$M_T$ region, $90 < M_T < 200$~GeV.
The result is $\Gamma_W = 2032 \pm 45_{\rm stat }\pm 57_{\rm syst}$~MeV, consistent
with the standard model expectation.

\end{abstract}
\maketitle
The decay widths of the $W$ and $Z$ bosons that mediate the weak
interaction are precisely predicted within the standard model (SM). At
Born level the $W$ width {\gammaw} and mass {\mw} are related through
the precisely determined Fermi coupling constant, {\Gf}. Beyond
leading order, higher-order electroweak (EW) and quantum-chromodynamic
(QCD) corrections, $\delta_{EW} \approx -0.4\%$ and $\delta_{QCD}
\approx 2.5\%$ respectively, modify the relation such that $\Gamma_W = \frac{3 G_F
M_W^3}{\sqrt{8}\pi}\left(1+\delta_{EW}+\delta_{QCD}\right)$~\cite{ref:ROSNER, ref:UNITS}.
The uncertainty on the SM prediction $\Gamma_W = 2091 \pm 2$~${\rm
MeV}$ is dominated by the uncertainty on {\mw} with smaller
contributions from the uncertainties on the higher order
corrections~\cite{ref:PDG}. 
Uncertainties on SM parameters, such as the Higgs boson mass, affect this
prediction very weakly,
and so a measurement allows an unambiguous test of the SM that can also be used
to constrain other SM parameters such as the $V_{cs}$ CKM matrix element~\cite{ref:RENTON}.
The average of previously published direct measurements of \gammaw\
from $p\overline{p}$ collisions at the
Tevatron~\cite{ref:RUN1_TEVATRON_COMBINATION} and $e^{+}e^{-}$
collisions at LEP-II~\cite{ref:LEP_WIDTHS} has a combined uncertainty
of 2.7\% with the most precise determination from a single experiment
(ALEPH) having an uncertainty of 5.1\%.  The most precise indirect
determination~\cite{ref:CDF_INDIRECT} of {\gammaw} from a measurement
of the ratio $R = \frac{\sigma(p\overline{p} \rightarrow W \rightarrow
\ell\nu)}{\sigma(p\overline{p} \rightarrow Z \rightarrow
\ell^+\ell^-)}$ has an uncertainty of 2\%.

This Letter presents the world's most precise direct determination of
{\gammaw} from a single experiment. The analysis uses $W
\rightarrow e\nu$ and $W \rightarrow \mu \nu$ data
with respective integrated luminosities of 370~pb$^{-1}$ and 330~pb$^{-1}$
collected by the CDF II detector at the Fermilab Tevatron.

Neutrinos are undetectable by the CDF II detector and hence the
invariant mass of the $W$ boson cannot be reconstructed. \gammaw\
is therefore determined from a fit to the distribution of
the $W$ transverse mass $M_T = \sqrt{2\left(p_{T}^{\ell} p_{T}^{\nu} -
\vec{p}_{T}^{\,\,\ell} \cdot \vec{p}_{T}^{\,\,\nu} \right) }$, where
{\ptlvec} and {\ptnuvec} are the measured transverse momentum of the
charged lepton and the transverse momentum of the neutrino as inferred 
from the observed missing transverse energy, respectively.
Events with \mt\ $>$ \mw\ arise predominantly from a combination of
non-zero $W$ width and finite detector resolution. The width component 
of the high-\mt\ line-shape falls off more slowly than the resolution 
component, allowing a precise \gammaw\ measurement from the \mt\ $>$ \mw\ 
events even in the presence of systematic uncertainties on the resolution.

The components of the CDF~II detector relevant to this analysis are
described briefly here; a more complete description can be found
elsewhere~\cite{ref:WMASS_PRD}. A silicon microstrip
detector~\cite{ref:SILICON} is used to measure the distance of closest
approach in the transverse plane, \ipar, of charged particles to the
beamline.  The momenta of charged particles are measured using a
96-layer open-cell drift chamber (COT)~\cite{ref:COT} inside a 1.4~T
solenoid.  Optimal $p_{T}$ resolution is obtained by constraining
trajectories to originate from the beamline. Electromagnetic and
hadronic calorimeters, arranged in a projective tower geometry, cover
the pseudorapidity range $|\eta| < 3.64$~\cite{ref:CDF_COORDS}. In the
region $|\eta| < 1.0$, a lead/scintillator electromagnetic calorimeter
(CEM)~\cite{ref:CEM} measures electron energies and proportional
chambers embedded at the shower maximum provide further information on
shower shapes and positions.  A system of drift chambers outside the 
calorimeters is used to identify muons in the region 
$|\eta| < 1.0$~\cite{ref:muon_detectors}.

$W \rightarrow e\nu$ candidate events are selected 
by a large transverse energy electron
trigger, and the electron shower is required to have transverse energy
$E_T^e > 25$~GeV~\cite{ref:CDF_COORDS} in the CEM. The ratio of the
energy measured in the CEM and the charged-track momentum measured in
the COT, $E/p$, must satisfy $0.8 < E/p < 1.3$. The ratio of energy
deposited in the hadronic (HAD) and CEM calorimeter towers is required
to satisfy $E_{\rm{HAD}}/E_{\rm{CEM}} < 0.07$. The electron shower
must be contained within a fiducial region of the CEM, away from
calorimeter cell boundaries, and must have 
a typical electron lateral shower profile on the projection
of the COT track in the CEM.  
Contamination by $Z \rightarrow e^+e^-$ events is reduced by rejecting 
events with an additional high $p_T$ track of opposite sign charge 
pointing to an un-instrumented region of the calorimeter.

$W \rightarrow \mu\nu$ candidate events are selected by a 
large $p_T$ muon trigger
and are required to contain a COT track, well matched to a
track segment in the muon chambers, with transverse
momentum \ptm\ $> 25$~GeV. The energy deposited in the
electromagnetic and hadronic calorimeters must be consistent with the
passage of a minimum-ionizing particle. Requirements on the track
\ipar\ and track fit $\chi^2$ are imposed to reject background.
Events consistent with cosmic rays or those with an
additional high-$p_{T}$ track consistent with $Z \rightarrow
\mu^+\mu^-$ decays are removed.

The existence of a neutrino is inferred from a transverse momentum
imbalance. The missing transverse momentum,
$\vec{p}_{T}^{\,\,\nu} \equiv -\left( \vec{p}_T^{\,\,\ell} + \vec{u}
\right)$, must satisfy \met~$>$ 25~GeV. 
The components of the recoil transverse energy vector $\vec{u}$ are defined as
$\sum_{i}E_i\sin\theta_i(\cos\phi_i,\sin\phi_i)$, for calorimeter towers $i$
with $|\eta| < 3.64$, excluding those traversed by and surrounding the charged
lepton. $\vec{u}$ receives contributions from initial-state
QCD radiation, underlying-event energy, final-state photon
radiation, and overlapping $p\overline{p}$ interactions.  
%
%
To reduce backgrounds and improve transverse mass resolution, the
recoil energy must satisfy $u < 20$~GeV. The $W \rightarrow e\nu~(W
\rightarrow \mu\nu)$ sample consists of 127432 (108808) candidate
events in the range $50 < M_T < 200$~GeV and 3436 (2619) in the high
\mt\ range of $90 < M_T < 200$~GeV.

Since the $W$ and $Z$ bosons share a common production mechanism and the momenta of $Z$ bosons can be 
directly reconstructed from their decay products, $Z \rightarrow \ell^+\ell^-$ decays are used
to model the detector's response to $W \rightarrow
\ell\nu$ events. Samples of $Z \rightarrow e^+e^-$ and $Z \rightarrow
\mu^+\mu^-$ candidates are selected by requiring two charged leptons, with the same
requirements as the $W$ lepton candidates, with 
the exception that the muon chamber track match requirement
is removed for one of the muons in the  $Z \rightarrow \mu^+\mu^-$ pair.
 The di-lepton invariant mass
is required to satisfy $80 < M^{\ell\ell} <
100$~GeV. Samples of 2909 $Z \rightarrow e^+e^-$ and 6271 $Z
\rightarrow \mu^+\mu^-$ events with recoil energy $u < 20$~GeV are
used to determine the scale and resolution of the lepton energy and
momentum measurements. 
A second set of $Z \rightarrow e^+e^-$ and $Z
\rightarrow \mu^+\mu^-$ control samples is defined with the $u$ cut
replaced by a di-lepton transverse momentum cut, $p_T^{\ell\ell} <
50$~GeV, in order to constrain the $W$ boson's transverse momentum spectrum
and to provide an empirical model of the recoil. 

%



The $W$ boson $M_{T}$ spectrum is modeled using a Monte-Carlo
simulation. The CTEQ6M~\cite{ref:CTEQ6M} parton distribution functions
(PDFs) are used, and $W$ boson invariant masses $\sqrt{\hat{s}}$ are
generated according to an energy-dependent Breit-Wigner distribution:
$\sigma(\hat{s}) \sim \left[\hat{s}\left(1-M_W^2/\hat{s} \right)^2 +
\hat{s}\Gamma_W^2/M_W^2\right]^{-1}$.
Higher order QCD effects are included by generating the $W$ bosons with a $p_T$
distribution from a NLO combined with resummation QCD calculation~\cite{ref:AK_ELLIS}
with the non-perturbative prescription of~\cite{ref:BNLY}.
Photon radiation from the charged lepton is simulated using
a $\cal{O}(\alpha)$ matrix-element calculation~\cite{ref:BERENDS_KLEISS}. 
Corrections for EW box diagrams 
are applied from the calculation of~\cite{ref:BAUR_BOX}.


The charged leptons and radiated photons are passed through a custom
detector simulation that models in detail the energy loss due to 
ionization and bremsstrahlung. 
The simulation also includes a parametric model of the $\vec{u}$
measurement as a function of the boson $p_T$, tuned on data as
described below.
The same
kinematic and geometric cuts used to select candidate events in the
data are applied to the simulation. 
The simulation produces $M_T$ spectra 
for \gammaw\ values ranging from 1.0 to 3.0 GeV 
with $M_W$ fixed at 80.403~GeV~\cite{ref:PDG}. 

This measurement relies on the accurate modeling of the $M_T$
distribution over a wide range. The most important sources of
systematic uncertainty affecting the $M_{T}$ shape arise from the
charged-lepton energy and momentum scales and resolutions, the recoil
modeling, and the presence of backgrounds. All systematic
uncertainties are evaluated by varying parameters in the simulation
and then fitting the resulting $M_T$ spectra with the
nominal spectra. Uncertainties have been calculated separately for the
fit region $M_T^{\rm cut} < M_T < 200$~GeV for $M_T^{\rm cut}$ values
ranging from 80 to 110~GeV.  While the statistical uncertainty
decreases as $M_T^{\rm cut}$ is lowered, the systematic uncertainty
increases. A value of $M_T^{\rm cut} = 90$~GeV gives the smallest
total uncertainty.  Backgrounds are added to the simulation $M_T$ spectra
which are then normalized to the
number of data events in the region $50 < M_T < 90$~GeV.

The COT momentum scale is determined from a fit to the $Z \rightarrow
\mu^+\mu^-$ invariant mass distribution with the $Z$ mass constrained
to the world average value~\cite{ref:PDG}. A consistent COT momentum scale is obtained
from fits to the invariant mass distributions of $J/\psi \rightarrow
\mu^+\mu^-$ and $\Upsilon \rightarrow \mu^+\mu^-$
events~\cite{ref:WMASS_PRD}. The difference between the three
determinations has a negligible effect on this analysis.
The contribution to the uncertainty on
$\Gamma_W$ in the $W \rightarrow \mu\nu$ channel
$\Delta\Gamma_W^{\mu\nu}$ arising from the 0.04\% uncertainty in the
COT momentum scale, due to the  $Z \rightarrow \mu^+\mu^-$ statistics, is 
$\Delta\Gamma_W^{\mu\nu} = 17$~MeV.

By scaling the resolutions predicted by a {\sc geant}~\cite{ref:GEANT}
simulation of the COT to match the observed di-muon invariant mass distribution in
$Z \rightarrow \mu^+\mu^-$ decays, we obtain a momentum resolution of
$\sigma(1/p_{T}) = (5.4 \pm 0.2 )\times 10^{-4}~\rm{GeV}^{-1}$. 
A consistent $\sigma(1/p_{T})$ is also determined using the $E/p$
distribution of the $W \rightarrow e\nu$ data. 
The combined uncertainties from the $Z \rightarrow \mu^+\mu^-$ and
$E/p$ fits for the COT resolution give $\Delta\Gamma_W^{\mu\nu}$ =
26~MeV.

The CEM energy scale and resolution are determined from fits to the
$Z\rightarrow e^+e^-$ invariant-mass distribution with the
$Z$ mass constrained to the world average value~\cite{ref:PDG} and to the $E/p$
distribution of electrons in $W \rightarrow e\nu$ events, using the calibration of $p$
described above.  The scales determined from the two methods are
consistent and are combined to form a weighted average with an uncertainty of 0.04\%. 
The contribution to the uncertainty on $\Gamma_W$ in the $W
\rightarrow e\nu$ channel $\Delta\Gamma_W^{e\nu}$ arising from this uncertainty
is $\Delta\Gamma_W^{e\nu} = 17$~MeV.

The CEM resolution fits constrain the constant term $\kappa$ in the
CEM resolution function $\sigma(E)/E = 13.5\%/\sqrt{E_T({\rm GeV})}
\oplus \kappa$~\cite{ref:CEM}.  
The $E/p$ and $Z\rightarrow e^+e^-$ fit results differ by 1.6 standard
deviations. They are combined and an uncertainty is assigned that spans both
values, as well as the values obtained when the $E/p$ fit region is
varied, to give $\kappa=1.1\pm0.4\%$.  
This uncertainty on the CEM
resolution gives $\Delta\Gamma_W^{e\nu} = 31$~MeV.

Energy loss by electrons and photons in the solenoid coil and
associated material prior to the CEM, as well as energy leakage into
the hadronic calorimeter, are parameterized based on the results of a
{\sc geant} simulation.
In addition to
these simulated sources of CEM non-linearity, an additional
per-particle intrinsic non-linearity is determined from the $W
\rightarrow e\nu$ and $Z \rightarrow e^+e^-$ data by fitting the $E/p$
distribution in bins of $E_T$. Its uncertainty
gives $\Delta\Gamma_W^{e\nu} = 12$~MeV, resulting
in a total uncertainty of 21~MeV on $\Gamma_W^{e\nu}$ from the
uncertainties on the electron energy scale determination.  Finally,
uncertainties in the modeling of very low-energy photons and 
the amount of passive
material prior to the COT give $\Delta\Gamma_W^{e\nu} = 13$~MeV.

The recoil transverse energy vector $\vec{u}$ is used to determine
$\vec{p}_{T}^{\,\,\nu}$ and hence $M_T$. 
Since $\vec{u}$ comes predominantly from initial-state QCD radiation,
which is balanced by the $W$ or $Z$ boson $p_{T}$,
we form an empirical model by parameterizing its response and resolution as a function of $p_T^{\ell\ell}$.
The parameters of the model are
varied according to the covariance matrices obtained in the fits to $Z$ data.
The resulting uncertainties on $\vec{u}$ from the recoil model give
$\Delta\Gamma_W^{e\nu}$ ($\Delta\Gamma_W^{\mu\nu}$) = 54~(49)~MeV.
The uncertainty in the modeling of the \ptw\ distribution is
determined by fitting the $p_T^{\ell\ell}$ distribution in $Z
\rightarrow \ell^+\ell^-$ decays and results in a 7~MeV common
uncertainty on \gammaw. 

Several background processes can mimic the $W$ signal. The process $W
\rightarrow \tau\nu \rightarrow \ell\nu\nu\nu$ has a signature similar to
$W \rightarrow \ell\nu$ decays, but with lower \mt. $Z\rightarrow
\ell^+\ell^-$ events, where only one lepton is identified, can be
reconstructed as $W$ candidates. These two backgrounds can be
accurately 
determined from Monte-Carlo simulation. QCD
multi-jet backgrounds arise when one jet mimics a charged lepton and
another is mismeasured to produce an energy imbalance.  
Since the region with apparently low \met\ is enriched in 
QCD background, the background normalization is estimated from a 
fit to the \met\ distribution in events where the \met\ and 
low \mt\ cut are not applied. 
The background \met\ spectrum is taken from data events in which some of
the charged-lepton identification cuts have been reversed, and the
signal spectrum is taken from the simulation.  A decay-in-flight
(DIF) background to the $W \rightarrow \mu\nu$ signal arises when
kaons or pions decay to $\mu\nu$ inside the COT, 
resulting in 
mismeasured muon momentum
and a large $\chi^2$ value
between the COT hits assigned to the track and the fitted track
trajectory. This background is estimated from a fit to the $\chi^2$
distribution.  The background spectrum is taken from events with a
large \ipar\ and the signal from $Z \rightarrow \mu^+\mu^-$ events,
which have negligible background. 
The background fractions over the entire region $50 < M_T < 200$~GeV 
are indicated in Figure~1.
In the $90 < M_T < 200$~GeV fit region the total background fraction
is $4.0 \pm 0.2\%$ ($10.8 \pm 0.3\%$) in the electron (muon) channel. 
Varying the background predictions within
these overall normalization uncertainties, as well as varying their $M_T$
shapes, causes variations in the electron (muon) $\Gamma_W$ of 32
(33)~MeV.  
The backgrounds that contribute the most to the width uncertainty
are the QCD multijet (DIF) backgrounds in the electron (muon) channel.


We also investigate small systematic uncertainties due to PDFs, $M_W$,
EW corrections, lepton identification, and acceptance.  The
uncertainty on \gammaw\ arising from PDFs is determined using the
variations defined by the CTEQ6M PDF eigenvector
basis~\cite{ref:CTEQ6M}.  The PDF error sets span a 90\% confidence
interval so the resulting \gammaw\ shifts are divided by 1.6 to obtain
1$\sigma$ uncertainties~\cite{ref:HUSTON}, giving an uncertainty of
16~MeV in both channels. A systematic uncertainty of 12~MeV is added in quadrature
to this to account for the effect of higher order QCD effects not
implemented in the Monte-Carlo simulation which was estimated from a
comparison of the width obtained using NLO and NNLO
PDFs~\cite{ref:MRST_NNLO}.  Varying $M_W$ by the uncertainty of
$\pm$29~MeV~\cite{ref:PDG} from the central value of 80.403 GeV
changes \gammaw\ by $\mp$9 MeV in each channel.

The impact of higher-order EW corrections is determined by comparing 
simulated samples of $W \rightarrow \ell\nu\gamma$ and $W \rightarrow \ell\nu\gamma\gamma$ 
events generated by {\sc photos}~\cite{ref:PHOTOS}.  Uncertainties on
$\Gamma_W^{e\nu}$ ($\Gamma_W^{\mu\nu}$) of 8~(1)~MeV were obtained. 
The correction due to EW box diagrams was determined to be 12~MeV in both channels. 
A systematic uncertainty of 6~MeV in the box diagram correction was
assigned from the dependence of the correction on the recoil
resolution.
%
%

The uncertainty in simulating lepton identification
variables was constrained from $Z \rightarrow \ell^+\ell^-$ decays and
results in a \gammaw\ uncertainty of 10~(6)~MeV in the electron (muon)
channel.  Variations in the simulation of the detector acceptance
results in a further small uncertainty of 3~(4)~MeV in the
electron (muon) channel. Table~\ref{table:TABLE_SYST} summarizes the
sources of uncertainty described above.

A binned likelihood fit to simulated $M_T$ spectra with $\Gamma_W$ as
a free parameter over
the region $90 < M_T < 200$~GeV gives $\Gamma_W = 2118 \pm
60_{\rm stat}$~MeV for the electron channel and $\Gamma_W = 1948 \pm
67_{\rm stat}$~MeV for the muon channel. Figure 1 shows the $M_T$
distributions of the data with the best fits.  The electron and
muon results have a common uncertainty of 27~MeV and are combined
using the BLUE method~\cite{ref:BLUE} to give 
$\Gamma_W = 2032 \pm 45_{\rm stat} \pm 57_{\rm syst}$~MeV. The combination has a $\chi^2$
of 1.6 and a total uncertainty of 73~MeV. No statistically significant difference is
found between fits using only positively and only negatively charged
leptons. As a cross-check, the $W$ width was also determined from a
fit to the charged-lepton transverse momentum, which has a different
sensitivity to many of the systematics, and a value of \gammaw\
consistent with the $M_T$ fit at the $<1 \sigma$ level was obtained.

The result presented in this Letter is the most precise direct
measurement of the $W$ width. It can be combined with published
Tevatron direct width
measurements~\cite{ref:RUN1_TEVATRON_COMBINATION} to give a hadron
collider average of $\Gamma_W = 2056 \pm 62$~MeV. A further
combinatioon with the preliminary value obtained from $e^+e^-$
collisions, $\Gamma_W = 2196 \pm 84$~MeV~\cite{ref:LEP2_WIDTH}, gives
a new world average value of $\Gamma_W = 2106 \pm 50$~MeV, in good
agreement with the SM prediction. 
%
%
\begin{table}[htbp]
\centering
\caption{\label{table:TABLE_SYST} The sources of uncertainty (in MeV) on
$\Gamma_W$ for the $W \rightarrow e\nu$ and $W \rightarrow \mu\nu$
measurements. If there is a correlated source of error between the
two measurements its contribution to each measurement is listed
in the third column, labeled C.}
\begin{tabular}{lccc}
\\
\hline 
\hline
Source         & $\Delta\Gamma_W^{e\nu}$ & $\Delta\Gamma_W^{\mu\nu}$ & C \\ \hline
Statistics                               &  60    &  67   &    \\ \hline
Lepton $E$ or $p$ scale                  &  21    &  17   &  12 \\
Lepton $E$ or $p$ resolution             &  31    &  26   &     \\
Electron energy loss simulation          &  13    &       &     \\
Recoil model                             &  54    &  49   &     \\
$p_T^W$                                  &   7    &   7   &   7 \\
Backgrounds                              &  32    &  33   &     \\
PDFs                                     &  20    &  20   &  20 \\
$M_W$                                    &   9    &   9   &   9 \\ 
EW radiative corrections                 &  10    &   6   &   6 \\
Lepton ID/acceptance                     &  10    &   7   &     \\   \hline
Total Syst.                              &  79    &  71   &  27 \\ \hline
Total (Stat. + Syst.)                    &  99    &  98   &  27 \\ \hline \hline
\end{tabular}
\end{table}
\begin{figure}
\includegraphics[totalheight=0.28\textheight]{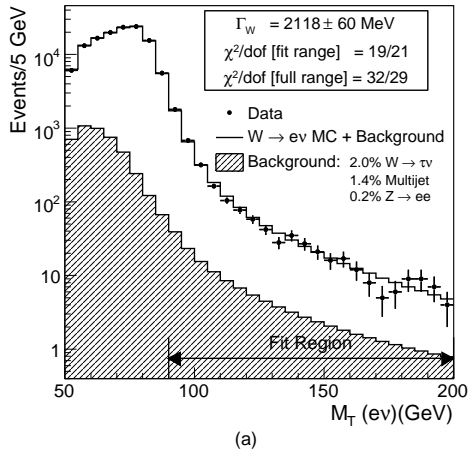}
\hfill
\includegraphics[totalheight=0.28\textheight]{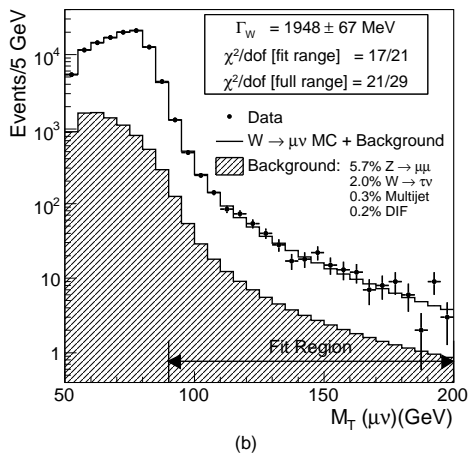}
\caption{\label{fig:MT_ELEC} The transverse mass distributions of the
$W \rightarrow e\nu$ data (a) and $W \rightarrow \mu\nu$  data (b)
compared to the best fit.}
\end{figure}

We thank the Fermilab staff and the technical staffs of the
participating institutions for their vital contributions. This work
was supported by the U.S. Department of Energy and National Science
Foundation; the Italian Istituto Nazionale di Fisica Nucleare; the
Ministry of Education, Culture, Sports, Science and Technology of
Japan; the Natural Sciences and Engineering Research Council of
Canada; the National Science Council of the Republic of China; the
Swiss National Science Foundation; the A.P. Sloan Foundation; the
Bundesministerium f\"ur Bildung und Forschung, Germany; the Korean
Science and Engineering Foundation and the Korean Research Foundation;
the Science and Technology Facilities Council and the Royal Society,
UK; the Institut National de Physique Nucleaire et Physique des
Particules/CNRS; the Russian Foundation for Basic Research; the
Comisi\'on Interministerial de Ciencia y Tecnolog\'{\i}a, Spain; the
European Community's Human Potential Programme; the Slovak R\&D Agency;
and the Academy of Finland.


\begin{thebibliography}{99}
\bibitem{ref:ROSNER} J. L. Rosner, M. P. Worah, and T. Takeuchi, Phys. Rev. D {\bf 49}, 1363 (1994).
\bibitem{ref:PDG} W.-M. Yao {\it et al.}, J. Phys. G {\bf 33}, 1 (2006).
\bibitem{ref:RENTON} P.~B.~Renton, Rept.\ Prog.\ Phys.\  {\bf 65}, 1271 (2002).
\bibitem{ref:RUN1_TEVATRON_COMBINATION} V. Abazov {\it et al.}, (CDF
Collaboration, D0 Collaboration, Tevatron EW Working
Group), Phys. Rev. D {\bf 70}, 092008 (2004).
\bibitem{ref:LEP_WIDTHS} S. Schael {\it et al.} (ALEPH
Collaboration), Eur. Phys. J. C {\bf 47}, 309 (2006); G. Abbiendi {\it et al.} (OPAL
Collaboration), {\it ibid.} {\bf 45}, 307 (2006); P. Achard {\it
et al.} (L3 Collaboration), {\it ibid.} {\bf 45}, 569 (2006);
P.Abreu {\it et al.} (DELPHI Collaboration), Phys. Lett. B {\bf 511}, 159 (2001).
\bibitem{ref:CDF_INDIRECT} D. Acosta {\it et al.} (CDF Collaboration),
Phys. Rev. Lett. {\bf 94}, 091803 (2005).
\bibitem{ref:WMASS_PRD} T.~Aaltonen {\it et al.}  (CDF Collaboration), arXiv:0708.3642 [hep-ex].
\bibitem{ref:SILICON} A. Sill {\it et al.}, 
Nucl. Instrum. Methods A {\bf 447}, 1 (2000); A. Affolder {\it et
al.}, {\it ibid.} {\bf 453}, 84 (2000).
\bibitem{ref:COT} A. Affolder {\it et al.}, 
Nucl. Instrum. Methods A {\bf 526}, 249 (2004).
\bibitem{ref:CDF_COORDS} Throughout this Letter units of $\hbar = c =
1$ are used such that momenta and masses are in units of MeV or
GeV. CDF uses a cylindrical coordinate system in which $\phi$ is the
azimuthal angle, $\theta$ is the polar angle, and $z$ points in the
proton beam direction.  The pseudorapidity, $\eta$, is defined as
$\eta \equiv -\ln\tan(\theta/2)$. The transverse plane is the
plane perpendicular to the $z$ axis. The transverse energy is defined
to be $E_T = E\sin\theta$.
\bibitem{ref:CEM} L. Balka {\it et al.}, Nucl. Instrum. Methods A {\bf 267}, 272 (1988).
\bibitem{ref:muon_detectors} G. Ascoli {\it et al.}, Nucl. Instrum. Methods A {\bf 268}, 33 (1998).
\bibitem{ref:CTEQ6M} J. Pumplin {\it et al.}, J. High Energy Phys. {\bf 0207}, 012 (2002).
\bibitem{ref:AK_ELLIS} P. B. Arnold and R. P. Kaufman, Nucl. Phys. {\bf
B349}, 381 (1991); P. B. Arnold and M. H. Reno, {\it ibid.} {\bf 319}, 37
(1989); P. B. Arnold, R. K. Ellis, and M. H. Reno, Phys. Rev. D {\bf 40}, 912
(1989).
\bibitem{ref:BNLY} F. Landry {\it et al.}, Phys. Rev. D {\bf 67}, 073016 (2003).
\bibitem{ref:BERENDS_KLEISS} F. A. Berends {\it et al.}, Z. Phys. C {\bf 27}, 155 (1985); F. A. Berends and
R. Kleiss, {\it ibid.} {\bf 27}, 365 (1985).
\bibitem{ref:BAUR_BOX}  U. Baur and D. Wackeroth, Phys. Rev. D {\bf 70}, 073015 (2004).
\bibitem{ref:GEANT} R. Brun and F. Carminati, CERN Program Library Long Writeup,
W5013, 1993 (unpublished), version 3.15.
\bibitem{ref:HUSTON} J. M. Campbell, J. W. Huston, and W. J. Stirling, Rep. Prog. Phys. {\bf 70}, 89 (2007). 
\bibitem{ref:MRST_NNLO} A.D. Martin, R.G. Roberts, W.J. Stirling, and R.S. Thorne, Phys. Lett. B {\bf 604}, 61 (2004).
\bibitem{ref:PHOTOS} E. Barberio and Z. Was,
Comput. Phys. Commun. {\bf 79}, 291 (1994); E. Barberio, B. van Eijk, and Z. Was,
{\it ibid.} {\bf 66}, 115 (1991).
\bibitem{ref:BLUE} L. Lyons, D. Gibaut, and P. Clifford,
Nucl. Instrum. Methods A {\bf 270}, 110 (1988).
\bibitem{ref:LEP2_WIDTH} J. Alcaraz {\it et al.} (LEP Electroweak Working Group), arXiv:hep-ex/0612034.

\end{thebibliography}
\end{document}